\documentclass[10pt, conference, letterpaper]{IEEEtran}
\usepackage{url}
\usepackage[utf8]{inputenc}
\usepackage{xcolor}
\usepackage{amsmath}

\DeclareMathOperator*{\argmin}{arg\,min}
\usepackage{multirow}
\usepackage{amssymb}
\usepackage{enumitem}
\usepackage{bm}
\usepackage{graphicx}

\IEEEoverridecommandlockouts

\usepackage{gensymb}

\usepackage[acronyms,nonumberlist,nopostdot,nomain,nogroupskip]{glossaries}
\usepackage{tablefootnote}
\usepackage{booktabs}
\usepackage{tabularx}
\usepackage{epsfig}
\usepackage[outdir=images_final/]{epstopdf}
\usepackage{tikz}
\usepackage{pgfplots}
\pgfplotsset{compat=newest} 
\pgfplotsset{plot coordinates/math parser=false} 
\newlength\fheight
\newlength\fwidth
\usetikzlibrary{plotmarks,patterns,decorations.pathreplacing,backgrounds,calc,arrows,arrows.meta,spy,matrix}
\usepgfplotslibrary{patchplots,groupplots}
\usepackage{tikzscale}
\usepackage{siunitx}

\usepackage{multirow}
\usepackage{algorithm2e, setspace}
\SetAlCapNameFnt{\footnotesize}
\SetAlCapFnt{\footnotesize}
\RestyleAlgo{ruled}
\SetKwComment{Comment}{/* }{ */}
\usepackage{algpseudocode}

\usepackage[font=scriptsize]{subcaption}
\usepackage[font=footnotesize]{caption}

\usepackage{mathtools}

\usepackage{dblfloatfix}    
\usepackage{colortbl}

\newacronym{3gpp}{3GPP}{3rd Generation Partnership Project}
\newacronym{adc}{ADC}{Analog to Digital Converter}
\newacronym{5g}{5G}{5th generation}
\newacronym{6g}{6G}{6th generation}
\newacronym{aimd}{AIMD}{Additive Increase Multiplicative Decrease}
\newacronym{am}{AM}{Acknowledged Mode}
\newacronym{amc}{AMC}{Adaptive Modulation and Coding}
\newacronym{aqm}{AQM}{Active Queue Management}
\newacronym{awgn}{AGWN}{Additive White Gaussian Noise}
\newacronym{balia}{BALIA}{Balanced Link Adaptation}
\newacronym{bdp}{BDP}{Bandwidth-Delay Product}
\newacronym{bf}{BF}{beamforming}
\newacronym{cc}{CC}{Congestion Control}
\newacronym{cdf}{CDF}{Cumulative Distribution Function}
\newacronym{cn}{CN}{Core Network}
\newacronym{cqi}{CQI}{Channel Quality Information}
\newacronym{cp}{CP}{Control Plane}
\newacronym{csirs}{CSI-RS}{Channel State Information - Reference Signal}
\newacronym{dc}{DC}{Dual Connectivity}
\newacronym{rb}{RB}{Resource Block}
\newacronym{dce}{DCE}{Direct Code Execution}
\newacronym{dci}{DCI}{Downlink Control Information}
\newacronym{udp}{UDP}{User Datagram Protocol}
\newacronym{dl}{DL}{downlink}
\newacronym{fcfs}{FCFS}{first-come-first-served}
\newacronym{dmr}{DMR}{Deadline Miss Ratio}
\newacronym{fspl}{FSPL}{free-space path loss}
\newacronym{dmrs}{DMRS}{DeModulation Reference Signal}
\newacronym{e2e}{E2E}{End-to-End}
\newacronym{ppp}{PPP}{Poission Point Process}
\newacronym{aoi}{AoI}{Area of Interest}
\newacronym{cpu}{CPU}{Central Processing Unit}
 \newacronym{gpu}{GPU}{Graphics Processing Unit}
 \newacronym{tpu}{TPU}{Tensor Processing Unit}
\newacronym{si}{SI}{Study Item}
\newacronym{ecn}{ECN}{Explicit Congestion Notification}
\newacronym{edf}{EDF}{Earliest Deadline First}
\newacronym{enb}{eNB}{eNodeB}
\newacronym{epc}{EPC}{Evolved Packet Core}
\newacronym{es}{ES}{Edge Server}
\newacronym{cav}{CAV}{Connected and Autonomous Vehicle}
\newacronym{fdma}{FDMA}{Frequency Division Multiple Access}
\newacronym{fdd}{FDD}{Frequency Division Duplexing}
\newacronym{upa}{UPA}{Uniform Planar Array}
\newacronym[firstplural=Radio Access Technologies (RATs)]{rat}{RAT}{Radio Access Technology}
\newacronym[firstplural=Radio Access Technology (RTs)]{rt}{RT}{Radio Technology}
\newacronym{fs}{FS}{Fast Switching}
\newacronym{isd}{ISD}{inter-site distance}
\newacronym{ftp}{FTP}{File Transfer Protocol}
\newacronym{gnb}{gNB}{Next Generation Node Base}
\newacronym{harq}{HARQ}{Hybrid Automatic Repeat reQuest}
\newacronym{hetnet}{HetNet}{Heterogeneous Network}
\newacronym{hh}{HH}{Hard Handover}
\newacronym{hol}{HOL}{Head-of-Line}
\newacronym{ia}{IA}{Initial Access}
\newacronym{imt}{IMT}{International Mobile Telecommunication}
\newacronym{iot}{IoT}{Internet of Things}
\newacronym{los}{LOS}{Line of Sight}
\newacronym{lte}{LTE}{Long Term Evolution}
\newacronym{m2m}{M2M}{Machine to Machine}
\newacronym{mac}{MAC}{Medium Access Control}
\newacronym{mc}{MC}{Multi-Connectivity}
\newacronym{mcs}{MCS}{Modulation and Coding Scheme}
\newacronym{mec}{MEC}{Mobile Edge Cloud}
\newacronym{mi}{MI}{Mutual Information}
\newacronym{mimo}{MIMO}{Multiple Input Multiple Output}
\newacronym{mmwave}{mmWave}{millimeter wave}
\newacronym{mptcp}{MPTCP}{Multipath TCP}
\newacronym{mr}{MR}{Maximum Rate}
\newacronym{mss}{MSS}{Maximum Segment Size}
\newacronym{mtd}{MTD}{Machine-Type Device}
\newacronym{mtu}{MTU}{Maximum Transmission Unit}
\newacronym{nfv}{NFV}{Network Function Virtualization}
\newacronym{vnf}{VNF}{ Virtualization Network Function}
\newacronym{gv}{GV}{ground vehicle}
\newacronym{vec}{VEC}{vehicular edge computing}
\newacronym{sdn}{SDN}{Software Defined Networking}
\newacronym{nlos}{NLOS}{Non Line of Sight}
\newacronym{nlosb}{NLOSb}{Building Non Line of Sight}
\newacronym{nlosv}{NLOSv}{Vehicle Non Line of Sight}
\newacronym{nr}{NR}{New Radio}
\newacronym{ofdm}{OFDM}{Orthogonal Frequency Division Multiplexing}
\newacronym{pdcch}{PDCCH}{Physical Downlonk Control Channel}
\newacronym{pdcp}{PDCP}{Packet Data Convergence Protocol}
\newacronym{pdsch}{PDSCH}{Physical Downlink Shared Channel}
\newacronym{pdu}{PDU}{Packet Data Unit}
\newacronym{pf}{PF}{Proportional Fair}
\newacronym{pgw}{PGW}{Packet Gateway}
\newacronym{phy}{PHY}{Physical}
\newacronym{pbch}{PBCH}{Physical Broadcast Channel}
\newacronym[plural=\gls{mme}s,firstplural=Mobility Management Entities (MMEs)]{mme}{MME}{Mobility Management Entity}
\newacronym{prb}{PRB}{Physical Resource Block}
\newacronym{pss}{PSS}{Primary Synchronization Signal}
\newacronym{pucch}{PUCCH}{Physical Uplink Control Channel}
\newacronym{pusch}{PUSCH}{Physical Uplink Shared Channel}
\newacronym{rach}{RACH}{Random Access Channel}
\newacronym{ran}{RAN}{Radio Access Network}
\newacronym{red}{RED}{Random Early Detection}
\newacronym{rf}{RF}{Radio Frequency}
\newacronym{rlc}{RLC}{Radio Link Control}
\newacronym{rlf}{RLF}{Radio Link Failure}
\newacronym{rrc}{RRC}{Radio Resource Control}
\newacronym{rrm}{RRM}{Radio Resource Management}
\newacronym{rr}{RR}{Round Robin}
\newacronym{rs}{RS}{Remote Server}
\newacronym{rsrp}{RSRP}{Reference Signal Received Power}
\newacronym{rss}{RSS}{Received Signal Strength}
\newacronym{rtt}{RTT}{Round Trip Time}
\newacronym{rw}{RW}{Receive Window}
\newacronym{rx}{RX}{Receiver}
\newacronym{sa}{SA}{standalone}
\newacronym{sack}{SACK}{Selective Acknowledgment}
\newacronym{sap}{SAP}{Service Access Point}
\newacronym{sch}{SCH}{Secondary Cell Handover}
\newacronym{scoot}{SCOOT}{Split Cycle Offset Optimization Technique}
\newacronym{sdma}{SDMA}{Spatial Division Multiple Access}
\newacronym{sinr}{SINR}{Signal to Interference plus Noise Ratio}
\newacronym{sm}{SM}{Saturation Mode}
\newacronym{snr}{SNR}{Signal to Noise Ratio}
\newacronym{son}{SON}{Self-Organizing Network}
\newacronym{ss}{SS}{Synchronization Signal}
\newacronym{srs}{SRS}{Sounding Reference Signal}
\newacronym{sss}{SSS}{Secondary Synchronization Signal}
\newacronym{tb}{TB}{Transport Block}
\newacronym{tcp}{TCP}{Transmission Control Protocol}
\newacronym{tdd}{TDD}{Time Division Duplexing}
\newacronym{tdma}{TDMA}{Time Division Multiple Access}
\newacronym{tfl}{TfL}{Transport for London}
\newacronym{tm}{TM}{Transparent Mode}
\newacronym{prr}{PRR}{Packet Reception Ratio}
\newacronym{trp}{TRP}{Transmitter Receiver Pair}
\newacronym{tti}{TTI}{Transmission Time Interval}
\newacronym{ttt}{TTT}{Time-to-Trigger}
\newacronym{tx}{TX}{Transmitter}
\newacronym{ue}{UE}{User Equipment}
\newacronym{ul}{UL}{uplink}
\newacronym{uml}{UML}{Unified Modeling Language}
\newacronym{um}{UM}{Unacknowledged Mode}
\newacronym{utc}{UTC}{Urban Traffic Control}
\newacronym{vm}{VM}{Virtual Machine}
\newacronym{rsrq}{RSRQ}{Reference Signal Received Quality}
\newacronym{rssi}{RSSI}{Received Signal Strength Indicator}
\newacronym{crs}{CRS}{Cell Reference Signal}
\newacronym{v2v}{V2V}{Vehicle-to-Vehicle}
\newacronym{v2i}{V2I}{Vehicle-to-Infrastructure}
\newacronym{v2n}{V2N}{Vehicle-to-Network}
\newacronym{v2x}{V2X}{Vehicle-to-Everything}
\newacronym{vn}{VN}{Vehicular Node}
\newacronym{dsrc}{DSRC}{Dedicated Short Range Communication}
\newacronym{ci}{CI}{context information}
\newacronym{voi}{VoI}{value of information}
\newacronym{gps}{GPS}{Global Positioning System}
\newacronym{qos}{QoS}{Quality of Service}
\newacronym{qoe}{QoE}{Quality of Experience}
\newacronym{ml}{ML}{Machine Learning}
\newacronym{ahp}{AHP}{Analytic Hierarchy Process}
\newacronym{lidar}{LIDAR}{Light Detection and Ranging}
\newacronym{sumo}{SUMO}{Simulation of Urban MObility}
\newacronym{wave}{WAVE}{Wireless Access in Vehicular Environment}
\newacronym{c-its}{C-ITS}{Connected Intelligent Transportation System}
\newacronym{dash}{DASH}{Dynamic Adaptive Streaming over HTTP}
\newacronym{http}{HTTP}{HyperText Transfer Protocol}
\newacronym{nt}{NT}{non-terrestrial}
\newacronym{ntc}{NTC}{non-terrestrial communication}
\newacronym{ntn}{NTN}{non-terrestrial network}
\newacronym{haps}{HAPS}{High Altitude Platform Station}
\newacronym{hap}{HAP}{High Altitude Platform}
\newacronym{leo}{LEO}{Low Earth Orbit}
\newacronym{meo}{MEO}{Medium Earth Orbit}
\newacronym{geo}{GEO}{Geostationary Earth Orbit}
\newacronym{uav}{UAV}{Unmanned Aerial Vehicle}
\newacronym{nsat}{nSAT}{Nanosatellite}
\newacronym{ehf}{EHF}{extremely high-frequency}
\newacronym{ioe}{IoE}{Internet of Everyone}
\newacronym{gan}{GaN}{Gallium Nitride}

\definecolor{steelblue}{RGB}{176,196,222}

\usepackage{hyperref}
\usepackage[capitalize]{cleveref}

\makeglossaries
\begin{document}
	
	\title{UAV/HAP-Assisted Vehicular Edge Computing \\ in 6G: Where and What to Offload?}
	
	\author{\IEEEauthorblockN{Alessandro Traspadini, Marco Giordani, Michele Zorzi}
	\IEEEauthorblockA{
		Department of Information Engineering, University of Padova, Italy. Email: \texttt{\{name.surname\}@dei.unipd.it}}}

	\maketitle
	
	\begin{abstract}
		In the context of \gls{6g} networks, \gls{vec} is emerging as a promising solution to let battery-powered ground vehicles with limited computing and storage resources offload processing tasks to more powerful devices.
		Given the dynamic vehicular environment, VEC systems need to be as flexible, intelligent, and adaptive as possible. 
		To this aim, in this paper we study the opportunity to realize VEC via \glspl{ntn}, where ground vehicles offload resource-hungry tasks to \glspl{uav}, \glspl{hap}, or a combination of the two. We define an optimization problem in which tasks are modeled as a Poisson arrival process, and apply queuing theory to find the optimal offloading factor in the system. Numerical results show that aerial-assisted VEC is feasible even in dense networks, provided that high-capacity HAP/UAV platforms are~available.
	\end{abstract}
	
\glsresetall

	\begin{tikzpicture}[remember picture,overlay]
	\node[anchor=north,yshift=-10pt] at (current page.north) {\parbox{\dimexpr\textwidth-\fboxsep-\fboxrule\relax}{
			\centering\footnotesize This paper has been submitted to IEEE for publication. Copyright may change without notice.}};
	\end{tikzpicture}
	
	\section{Introduction}
	\label{sec:intro}
	\Glspl{ntn} are expected to be a key component of \gls{6g} networks, as a means to provide cost-effective and high-capacity connectivity via \glspl{uav}, \glspl{hap}, and satellites~\cite{giordani2021non}. For example, \glspl{ntn} can promote  on-demand service continuity  when terrestrial towers are out of service (e.g., in emergency situations), and complement terrestrial networks in remote areas, hence representing a promising technology in combating the digital divide~\cite{Chaoub20216g}.

	Recently, \glspl{ntn} have been envisioned to provide 3D flexible coverage and enhanced road safety in \gls{v2x} networks~\cite{shinde2021towards}. 
	In this context, the success of these networks relies on the availability of massive data from on-board sensors, to guarantee accurate perception of the environment~\cite{zhang2018vehicular}. However, data processing based on machine learning (from compression~\cite{nardo2022point} to object detection and recognition~\cite{rossi2021role}, from tracking and trajectory prediction to data dissemination~\cite{baek2020vehicle}) requires extensive computational resources, which may be challenging for current vehicular terminals~\cite{v2x_offloading}.

	To solve this issue, the research community is investigating \gls{vec}, aiming at offloading resource-hungry computation, caching, and/or storage tasks to more powerful edge/distributed servers~\cite{liu2021vehicular}.	
	While implementing VEC on terrestrial roadside infrastructures may be costly and inefficient~\cite{v2x_uav}, aerial-assisted VEC is seen as a valuable solution to satisfy 6G V2X requirements.
	Equipped with a computing server, \glspl{uav} can support ubiquitous broadband connectivity in favorable \gls{los} conditions, and the physical proximity between the computing aerial node and the ground can promote low latency, reliable privacy protection, and high-resolution context awareness. 
	For example  the authors in~\cite{uav_position} investigated a UAV-enabled offloading system,  where the positions of UAVs were optimized to minimize resource consumption, while Hayat \emph{et al.,} in~\cite{hayat2021edge}, studied edge computing for drone navigation. 
	Lakew \emph{et al.,} in~\cite{hap_offloading}, explored the opportunity of data offloading via \glspl{hap}, in light of their payload capability, stable deployment in the stratosphere, and large coverage footprint. The potential of HAP-enabled edge computing has been also demonstrated in areas without ground network coverage like countrysides and mountains~\cite{ke2021edge}. 
	The availability of multi-layered hierarchical networks~\cite{wang2020potential}, i.e., the orchestration among UAV and HAP platforms, could further enhance \gls{vec} performance, and offer higher resilience and flexibility compared to standalone deployments.
	However, the limited communication range and the inherent latency associated with aerial networks may jeopardize \gls{vec} offloading, thereby making the design of these systems not trivial. 

	Along these lines, in this paper we validate the \gls{vec} paradigm for 6G networks, thus we investigate a \gls{v2x} scenario in which \glspl{gv} offload intensive computing tasks to aerial nodes with rich availability of resources.
	To do so, we formalize an optimization problem to minimize the processing time under latency and computational capacity constraints.
	The VEC system is characterized as a set of queues in which tasks are modeled according to a Poisson arrival process, at a rate proportional to the number of \glspl{gv}.
	Compared to literature papers, we analyze both \gls{uav} and \gls{hap} layers, and a combination of the two, as a function of several V2X-specific parameters, including the density of GVs and the size of the transmitted data. 
	For the first time, we shed light on the optimal computing capacity of \glspl{gv}/\glspl{uav}/\glspl{hap} that guarantees that processing operations never exceed the inter-frame time of the sensors as a means to support real-time V2X services.
	We demonstrate that there exists an optimal offloading factor resulting in low computational burden on board of vehicles and communication demands.
	Specifically, HAP-assisted VEC is more desirable than UAV-assisted VEC, which is limited by hardware and energy constraints.

	The rest of the paper is organized as follows. In Sec.~\ref{sec:system_model} we introduce our system model, in Sec.~\ref{sec:optimizing_offloading_for_ntns} we present our offloading algorithms, in Sec.~\ref{sec:performance_evaluation} we discuss our results, whereas conclusions are summarized in Sec.~\ref{sec:conclusions_and_future_works}.

	\section{System Model} 
	\label{sec:system_model}

		\begin{figure}[t!] 
		\centering
		\setlength{\belowcaptionskip}{-0.5cm}
		\includegraphics[width=0.9\columnwidth]{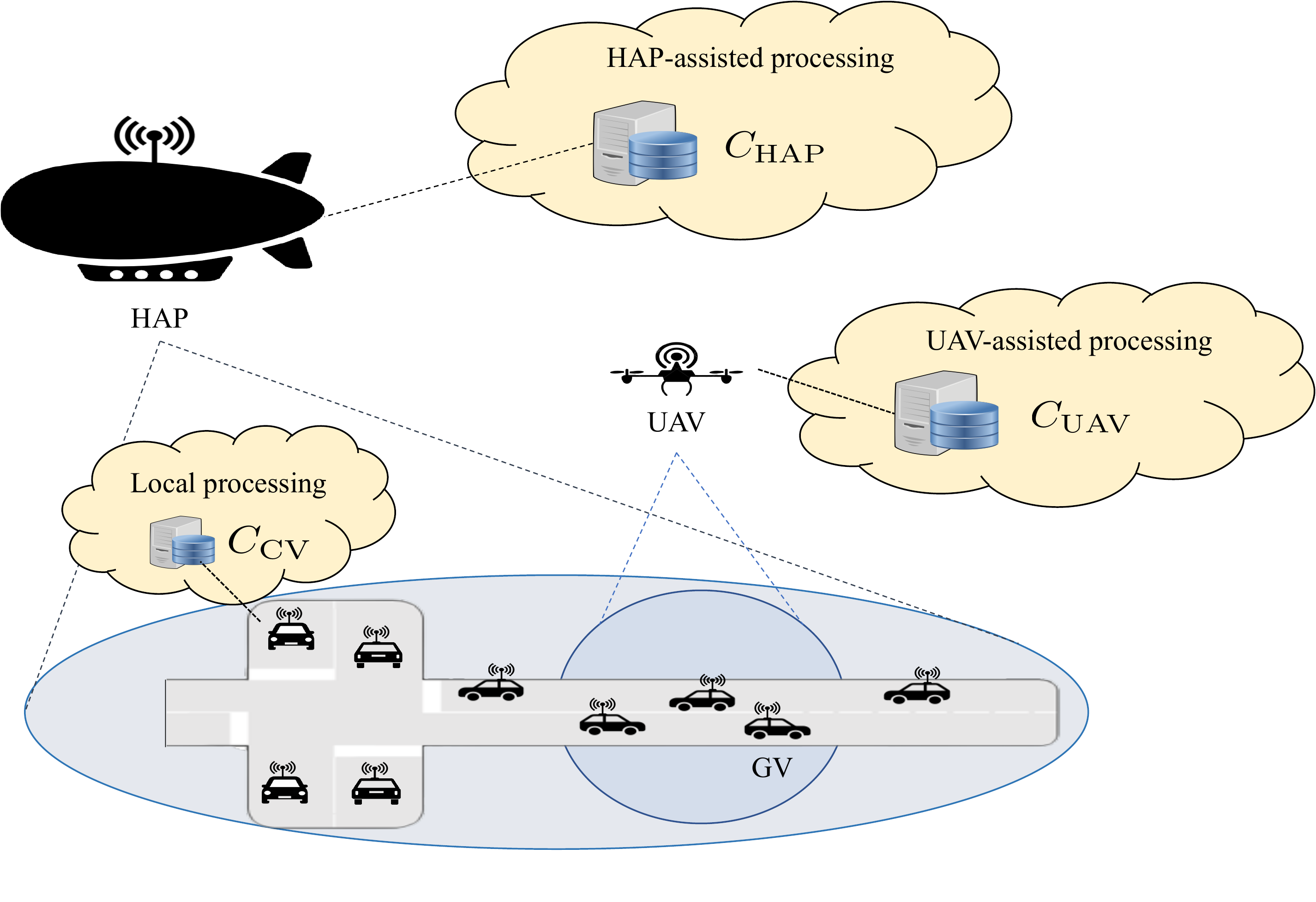}
		\caption{System model for NTN-assisted VEC.}
		\label{img:scenario}
		\centering
	\end{figure}

	In this section we present our research problem (Sec.~\ref{sub:formulation}), as well as the delay (Sec.~\ref{sub:delay_model}), channel (Sec.~\ref{sub:channel_model}) and queuing (Sec.~\ref{sub:queuing_model}) models.

	\subsection{Problem Formulation} 
	\label{sub:formulation}

	We consider a \gls{v2x} scenario at a road intersection in which $k$ GV/km$^2$ are deployed over an \gls{aoi} of $A$~$\text{km}^2$.
	The system may be overseen by HAP and/or UAV platforms flying over the \gls{aoi}; we assume that $A$ is smaller than the coverage footprint shaped by both UAVs and HAPs, as illustrated in Fig.~\ref{img:scenario}.

	We analyze the problem of \emph{cooperative perception} where each GV processes sensor's perceptions of $n_{\rm UL}$~bits, generated at a constant rate $r$, to identify road objects and make intersection crossing decisions accordingly~\cite{higuchi2019value}. 
	Each perception involves a constant computational load ${C}$ for detecting and tracking entities at the intersection~\cite{rossi2021role}.
	We assume that a fraction ${(1-\eta) C}$ of the computational load is processed on board of the GV (local processing), while the remaining computational load $\eta C$ is offloaded to NTN servers with high computation and storage capacity (VEC-assisted processing).
	The optimal offloading factor $\eta^*\in[0,1]$ must be dimensioned so as to minimize the total processing time.

	In the first case, data processing is performed by the GV, offering a limited computation capacity $C_{\rm GV}$ due to the high price of powerful \glspl{cpu} on board of current low-budget car models.
	In the second case, data processing is executed at the HAP/UAV's side.
	While resulting in low computational burden on board, transferring information over a wireless link and processing data at a remote device introduces a non-negligible delay. 
	Despite size, weight, and power limitations, high-end UAVs can be equipped with high-performance computing facilities, and leverage stronger computation capacity than available on board, i.e., $C_{\text{UAV}}\geq C_{\text{GV}}$~\cite{v2x_uav}.
	In addition, HAPs can be operated by solar panels which provide efficient and continuous energy resources, and their large payload permits the installation of high-performance platforms like \gls{gpu} and \gls{tpu}, to further speed up the computing tasks. As such, the computation capacity of HAPs is $C_{\text{HAP}}\geq C_{\text{UAV}}\geq C_{\text{GV}}$.

	\subsection{Delay Model} 
	\label{sub:delay_model}
	In this section we evaluate the processing time in case of local or VEC-assisted processing.

	\paragraph{Local processing} 
	The processing time does not involve data offloading to edge servers, and is simply given by:
	\begin{equation}
		t_{\rm LP}(\eta) =\left[(1-\eta){C}\right]/{C_{\rm GV}}.
	\end{equation}

	\paragraph{VEC-assisted processing} 
	Data offloading requires the wireless transfer of a sensor's perception of size $n_{\rm UL}$ produced on board to NTN servers, and the processed output of size $n_{\rm DL}\leq n_{\rm UL}$ to be sent back to the GV, which results into a capture-to-output delay given by:
	\begin{equation}
		t_{\rm VEC}(\eta)=2\tau_p+t_{\rm UL}+t_{\rm DL} + t_{\rm p,VEC}(\eta),
		\label{eq:t_vec}
	\end{equation}
	where $\tau_p$ is the propagation delay, $t_{\rm UL}$ and $t_{\rm DL}$ are the~transmissions delays in \gls{ul} and \gls{dl} from the GV to the NTN servers and vice versa, respectively, and $t_{\rm p,VEC}$ is the processing time at the NTN servers, which is studied in Sec.~\ref{sub:queuing_model} by analyzing the queuing model.

	In Eq.~\eqref{eq:t_vec}, $\tau_p$ depends on the distance $d$ between the generic GV and the NTN device. Assuming, without loss of generality, that GVs are uniformly distributed over the AoI $A$, and that the NTN device is placed right above the center of $A$ at an altitude $h_0$, we can write the average distance $\bar{d}$ as
	\begin{equation}
	\bar{d} = \sqrt{\frac{A}{2\pi}+h_0^2}.
	\end{equation}
	Thus, the average propagation delay is $\tau_p = \bar{d}/c_0$,
	where $c_0$ is the speed of light.
	The transmission delays, instead, are a function of the capacity of the \gls{ul} ($R_{\rm{UL}})$ and \gls{dl} ($R_{\rm{DL}}$) channels, and the number of vehicles in the system, i.e., 
	\begin{equation}
		\label{eq:txdelays}
			t_{\rm \ell} = \left(k\cdot A \cdot n_{\rm \ell}\right)/R_{\rm{\ell}},\quad \ell \in\{\text{UL},\,\text{DL}\}.
	\end{equation}

	\subsection{Channel Model} 
	\label{sub:channel_model}
	
	VEC involves offloading data to/from edge servers for processing. According to the 3GPP specifications~\cite{38821}, high-capacity transmissions can be established if NTN platforms operate at \glspl{mmwave}, where the huge bandwidths available may offer the opportunity of ultra-fast connections via highly directional antennas~\cite{giordani2020satellite}.
	According to the 3GPP, the \gls{snr} between transmitter $i$ and receiver $j$ can be computed as 
	\begin{equation}
		\label{eq:snr}
		\gamma_{ij} = \text{EIRP}_i + ({G_j}/{T}) - \text{PL}_{ij} - k - B,
	\end{equation}
	where the EIRP is the effective isotropic radiated power, ${G_j}/{T}$ is the receive antenna-gain-to-noise-temperature, PL is the path loss, $k$ is the Boltzmann constant, and $B$ is the bandwidth.
	The EIRP depends on the transmit antenna gain ($P_{T}$) and power ($G_{T}$) and the cable loss ($L_C$), and is given by~\cite{38821}
	\begin{equation}
		\text{EIRP}_i =  P_{T_i} - L_{C} + G_{T_i}.
	\end{equation}
	In turn, the antenna-gain-to-noise-temperature depends on the characteristics of the receiver, and can be computed as
	\begin{equation}
		\frac{G_j}{T} \resizebox{.85\hsize}{!}{$=  G_{R_j} - N_{f_j}-10\log_{10}\left[T_{0_j} + \left(T_{a_j} - T_{0_j} \right)\cdot 10^{\frac{-N_{f_j}}{10}}\right]$}
	\end{equation}
	where $G_R$ is the receive antenna gain, $N_f$ is the noise figure, and $T_0$ and $T_a$ are the ambient and antenna temperature.
	
	For the altitude range at which UAVs typically fly (below 500 m), the channel does not involve complete penetration through the atmosphere. Therefore, a simple \gls{fspl} model can be considered, and we have that 
	\begin{equation}
		\text{PL} = \text{FSPL} = 92.45 + 20\log_{10}(f_c) + 20\log_{10}(\bar{d}),
	\end{equation}
	where $f_c$ is the carrier frequency expressed in GHz, and $\bar{d}$ is the average distance between the transmitter and the receiver expressed in km.
	For the HAP channel, instead, the signal in the \gls{mmwave} bands undergoes several stages of attenuation through the atmosphere, in particular scintillation loss (PL$_{s}$, due to sudden changes in the refractive index caused by variation of the temperature, water vapor content, and barometric pressure) and atmospheric absorption loss (PL$_{g}$, due to dry air and water vapor attenuation).	Hence, the total path loss for \glspl{hap} can be expressed as
	\begin{equation}
	\text{PL} =  \text{FSPL} + \text{PL}_{g} + \text{PL}_{s}.
	\end{equation}
	For a more complete description of the channel model in NTN scenarios, we refer the interested reader to~\cite{wang2020potential}.
	
	Based on the channel model described above, we have that the average ergodic (Shannon) capacity $R_\ell$ shown in Eq.~\eqref{eq:txdelays}, i.e., the maximum achievable data rate, is given by:
	\begin{equation}
	R_{\ell} =  B \log_{2}(1+10^{\frac{\gamma_{ij}}{10}}), \quad \ell \in\{\text{UL},\,\text{DL}\}.
	\end{equation}

	\subsection{Queuing Model} 
	\label{sub:queuing_model}
	Data offloading involves a deterministic amount of time to be resolved, equal to the ratio between the fraction $\eta C$ of the computation load that is processed by the \gls{ntn} node and its computation capacity.
	This system can be modeled as an M/D/$c$ queue, in which arrivals of offloaded tasks are distributed as a Poisson process, and they are served following a \gls{fcfs} discipline.
	The number of servers $c$ is equal to the number of parallel processes that \gls{ntn} devices can handle, and there is only one waiting line (i.e., single channel queue).
	
	Based on the approximation introduced in \cite{queueinganalyzer}, the average waiting time in an M/D/$c$ queue is given by
	\begin{equation}
		W_q{({\text{M/D/}c})} \approx \frac{1}{2} W_q{({\text{M/M/}c})},
		\label{eq:w_q}
	\end{equation}
	where $W_q{{({\text{M/M/}c})}}$ denotes the waiting queuing time for an M/M/$c$ queue, which can be derived from the average length of the M/M/$c$ queue, $L_q{({\text{M/M/}c})}$, using Little's Law~\cite{benvenuto_zorzi}, i.e.,
	\begin{equation}
	W_q{({\text{M/M/}c})} = {L_q{({\text{M/M/}c})}}/{\lambda},
	\end{equation}
	where $\lambda= r \cdot k \cdot A$ is the arrival rate. $L_q{({\text{M/M/}c})}$ is computed~as
	\begin{equation}
	L_q{({\text{M/M/}c})} = \frac{G}{c-G} \cdot \mathcal{C}(c,G),
	\end{equation}
	where $G$ is the offered traffic, i.e., the ratio between the arrival rate and the service rate (that is $\mu = {C_i}/{(\eta C)}$, for $i\in\{\text{UAV, HAP}\}$), and $\mathcal{C}(c,G)$ is the probability that all the servers are busy and it is obtained as
	\begin{equation}
	\mathcal{C}(c,G) = \frac{1}{\sum_{r=0}^{c-1}\frac{G^r}{r!}+\frac{c \cdot G^c}{c! (c-G)}} \cdot \frac{c \cdot G^c}{c!(c-G)}.
	\end{equation}
	
	From the waiting time in Eq.~\eqref{eq:w_q}, we have that the processing time for offloading at the VEC servers shown in Eq.~\eqref{eq:t_vec} can be computed, for $i\in\{\text{UAV, HAP}\}$, as
	\begin{equation}
	t_{p,\text{VEC}}(\eta)= W_q{(\text{M/D}/c)} + ({1}/{\mu})= W_q{(\text{M/D}/c)} + \frac{\eta C}{C_i}.
	\end{equation}
	
	\section{Optimal Offloading for NTN-Assisted VEC} 
	\label{sec:optimizing_offloading_for_ntns}
	In this section we formalize an optimization problem to identify the optimal offloading factor $\eta^*$ to minimize the total processing time.
	In case of VEC-assisted processing, we distinguish between standalone (Sec.~\ref{sub:standalone}) and hybrid (Sec.~\ref{sub:hybrid}) offloading, depending on whether or not UAVs' and HAPs' computing facilities are used in combination.

	\subsection{Standalone Offloading}	
	\label{sub:standalone}

	In this case, a fraction $\eta^*_i$, $i\in\{\text{UAV, HAP}\}$, of the computational load $C$ is offloaded to either UAV or HAP platforms, that operate individually.
	The optimization problem in this configuration can be written as:
	\begin{subequations}
		\begin{alignat}{2}
		&\argmin_{\eta_i} &\quad& \max\left[t_\text{LP}(\eta_i),t_\text{\rm VEC}(\eta_i)\right], \notag\\
		&\text{subject to} & & G(\eta_i)< c, \label{eq:queue_constraint}\\
		&  & & \eta_i \in [0,1].
		\end{alignat}
	\end{subequations}
	Constraint ~\eqref{eq:queue_constraint} is necessary to keep the queue stable, which gives the following upper bound for $\eta_i$, i.e., 
	\begin{equation}
	\label{eq:bound}
	G(\eta_i)< c \: \Rightarrow \:  \eta_{i,\rm max} = \frac{c \cdot C_i}{C \cdot k \cdot A \cdot r}.
	\end{equation}

	The problem can be solved following an iterative procedure.
	If $t_\text{\rm LP}(\eta_i) < t_\text{\rm VEC}(\eta_i)$, data offloading is not convenient compared to local processing, and eventually results in an additional overhead  due to data transmission to/from NTN platforms. In this condition, the optimal solution would be to set $\eta_i^*=0$ (fully-local processing).
	Otherwise, it can be shown that the optimal solution of the problem is
	\begin{equation}
		\label{eq:opt_standalone}
		t_\text{LP}(\eta_i^*) = t_\text{VEC}(\eta_i^*).
	\end{equation}
	For simplicity, the algorithm stops when the difference between $t_\text{LP}$ and $t_\text{VEC}$ satisfies the following condition:
	\begin{equation}
		|t_\text{LP}(\ell) - t_\text{VEC}(\ell)| < \xi  \cdot  t_\text{VEC}(\ell).
	\end{equation}
	 The correctness of the algorithm is given by the fact that $t_\text{LP}$ and $t_\text{VEC}$ are, respectively, monotonically increasing and monotonically decreasing functions.

	\subsection{Hybrid Offloading}	
	\label{sub:hybrid}
	Initial studies have demonstrated that the availability of multi-layered hierarchical networks represents a promising technology to solve coverage and latency constraints in NTN scenarios~\cite{wang2020potential}.
	Therefore we analyzed the case in which GVs can offload data to both \gls{uav} and \gls{hap} simultaneously.
	The optimization problem is described as follows:
	\begin{subequations}
		\begin{alignat}{2}
		&\argmin_{\eta_\text{UAV},\eta_\text{HAP}} & \,& \max\left[t_\text{VEC}(\eta_\text{UAV}),t_\text{VEC}(\eta_\text{HAP}),
		t_\text{LP}(\eta_\text{UAV} + \eta_\text{HAP}) \right],\notag\\
		&\text{subject to} & & G(\eta_\text{UAV})< c_\text{UAV}, \: G(\eta_\text{HAP})< c_\text{HAP},  \label{constraint_queue_HAP}\\
		& & & \eta_\text{HAP},\eta_\text{UAV},\: \eta_\text{HAP}+\eta_\text{UAV}  \in [0,1].
		\end{alignat}
	\end{subequations}
	Constraints~\eqref{constraint_queue_HAP} are derived from Eq.~\eqref{eq:bound}, and are necessary to keep both \gls{uav} and \gls{hap} queues stable,  where $G(\eta_\text{UAV})$ and $G(\eta_\text{HAP})$ are the offered traffic, and $c_\text{UAV}$ and $c_\text{HAP}$ are the number of servers of UAVs or HAPs, respectively.

	The problem can be solved following a two-step procedure.
	\begin{itemize}
		\item \emph{Step 1:} Identify the optimal offloading factor $\eta^*_{\rm UAV}$ for the computational load $C$ in case of standalone UAV offloading. This problem is solved through the procedure described in Sec.~\ref{sub:standalone} for standalone UAV offloading, which requires a total processing time $t^*_{\rm SO, UAV}$.
		\item \emph{Step 2:} Evaluate whether HAPs can reduce the total processing time $t^*_{\rm SO, UAV}$. If $t^*_{\rm SO, UAV}<t_\text{\rm VEC}(\eta_{\rm HAP})$, HAP-assisted offloading is not convenient compared to standalone UAV offloading. In this condition, the optimal solution would be to set $\eta^*_{\rm HAP}=0$.
		Otherwise, hybrid offloading is desirable, and it can be shown that the optimal solution of the problem~is
	\begin{equation}
		\label{eq:opt_hybrid}
		t_\text{VEC}(\eta^*_{\rm HAP}) = t^*_{\rm SO, UAV}.
	\end{equation}
	For simplicity, the algorithm stops when we have:
	\begin{equation}
			|t_\text{VEC}(\ell_{\rm HAP}) - t^*_{\rm SO, UAV}|< \xi  \cdot  t_\text{VEC}(\ell_{\rm HAP})
	\end{equation}

	\end{itemize}

	\section{Performance Evaluation} 
	\label{sec:performance_evaluation}

	In this section, after introducing our simulation setup, we evaluate the performance of the proposed offloading schemes in different V2X scenarios.

	\subsection{Simulation Setup and Parameters} 
	\label{sub:simulation_setup_and_parameters}

\def\arraystretch{1.2}
\begin{table}[!t]
\centering
\footnotesize
\caption{System parameters.}
\label{tab:params_table}
\begin{tabular}{l|c|c|c}
\hline
\textbf{Parameter} & \textbf{GV} & \textbf{UAV} & \textbf{HAP}  \\
\hline
Sensor's perception rate ($r$) [fps] & \multicolumn{3}{c}{10}\\\hline
Sensor's perception size ($n_{\rm UL}$) [Mb] & \multicolumn{3}{c}{1}\\\hline
Processed data size ($n_{\rm DL}$) [Mb] & \multicolumn{3}{c}{0.1}\\\hline
Computational load ($C$) [GFLOP] & \multicolumn{3}{c}{100}\\\hline
Density of GVs ($k$) [GV/km$^2$] & \multicolumn{3}{c}{[25,500]}\\\hline
Area of interest ($A$) [km$^2$] & \multicolumn{3}{c}{1}\\\hline
Carrier frequency ($f_c$) [GHz] & \multicolumn{3}{c}{38}\\\hline
Bandwidth ($B$) [MHz] & \multicolumn{3}{c}{400}\\\hline
Altitude ($h_0$) [km]  & 0 & 0.1 & 20 \\\hline
Path loss (PL) [dB] & N/A & 101.98 & 172.76 \\\hline
EIRP [dBW] & 29 & $-10$ & 27.9 \\\hline
Rx. gain/noise temperature ($G/T$) [dB/K] & 12.15 & $-11.6$ & 27.7 \\\hline
Computational capacity ($C_i$) [GFLOP/s] & 500 & 1500 & 3500 \\\hline
Maximum number of servers ($c_i$) & 1 & 4 & 12 \\\hline
\end{tabular}
\vspace{-0.33cm}
\end{table}

	Simulation parameters, if not otherwise specified, are reported in Table~\ref{tab:params_table}.
	Our scenario consists of an area of size $A=1$ km$^2$, where $k$ GV/km$^2$ are deployed according to a \gls{ppp}.
	All devices operate at $f_c =$ 38~GHz, with a bandwidth of $B=400$~MHz.
	
	Each GV produces a sensor's perception (e.g.,  an RGB camera image) of size $n_\text{UL}=1$~Mb, at a constant rate $r=10$~fps.
	Each perception involves a constant computational load $C=100$~GFLOP for V2X-related processing (e.g., object detection and classification), and the processed output (e.g., bounding boxes) is eventually returned to the GVs in a packet of size $n_\text{DL} = 0.1$~Mb.
	GVs, UAVs and HAPs have a computational capacity of $C_{\rm GV}=500$, $C_{\rm UAV}=1500$ and $C_{\rm HAP}=3000$ GFLOP/s, respectively. 
	However, as the performance of computing hardware continues to hit new milestones, it is expected that connected/automated cars, as well as UAV/HAP platforms, will have more computing power in the future. 
	To capture this, numerical results will be given considering a range of values for the computational capacity, to simulate different computing conditions.

	The performance of the offloading schemes in Sec.~\ref{sec:optimizing_offloading_for_ntns} are evaluated in terms of the total time it takes to process the computational load $C$, as a function of the density of GVs, the size of the transmitted data, as well as the processing capability of the nodes.
	Ideally, the processing time should be lower than the frame rate $r$ of the sensors (100 ms in our scenario), so as to ensure real-time driving responses.
	
	\subsection{Numerical Results} 
	\label{sub:numerical_results}
	\subsubsection{Users density}

\begin{figure}[t!]
		\centering
		\subfloat[][$C_{\rm GV}= 200$~GFLOP/s.]
		{	\label{fig:users_density_b}
			\includegraphics[width=.87\columnwidth]{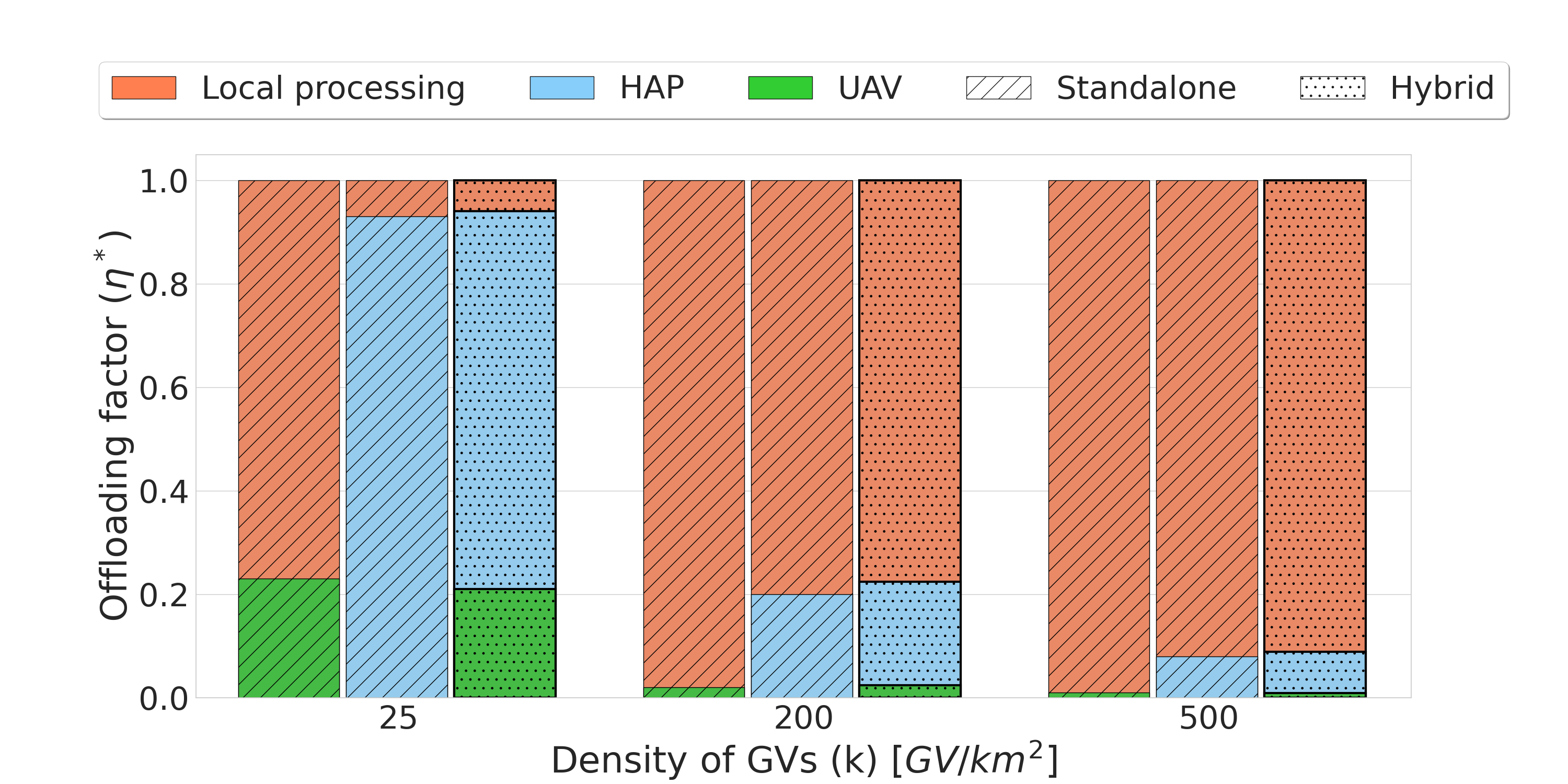}} \\
			\subfloat[][$C_{\rm GV}= 1000$~GFLOP/s.]
			{	\label{fig:users_density_d}
			\includegraphics[width=.87\columnwidth]{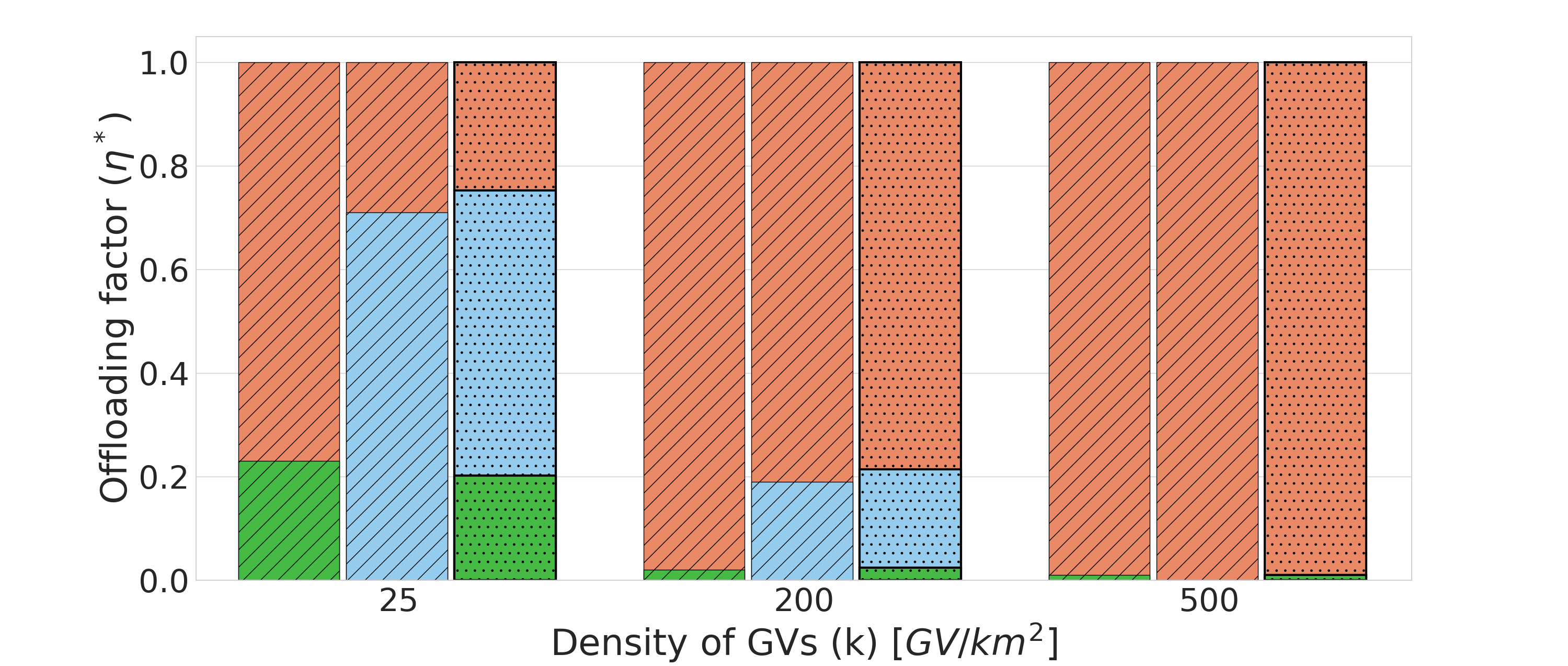}}\\
		\caption{Total processing time for different offloading schemes, vs. $C_{\rm GV}$ and~$k$.\vspace{-0.6cm}}
		\label{fig:users_density_time}
	\end{figure}

\begin{figure}[t!]
		\centering
			\subfloat[][$C_{\rm GV}= 200$~GFLOP/s.]
		{	\label{fig:users_density_a}
			\includegraphics[width=.87\columnwidth]{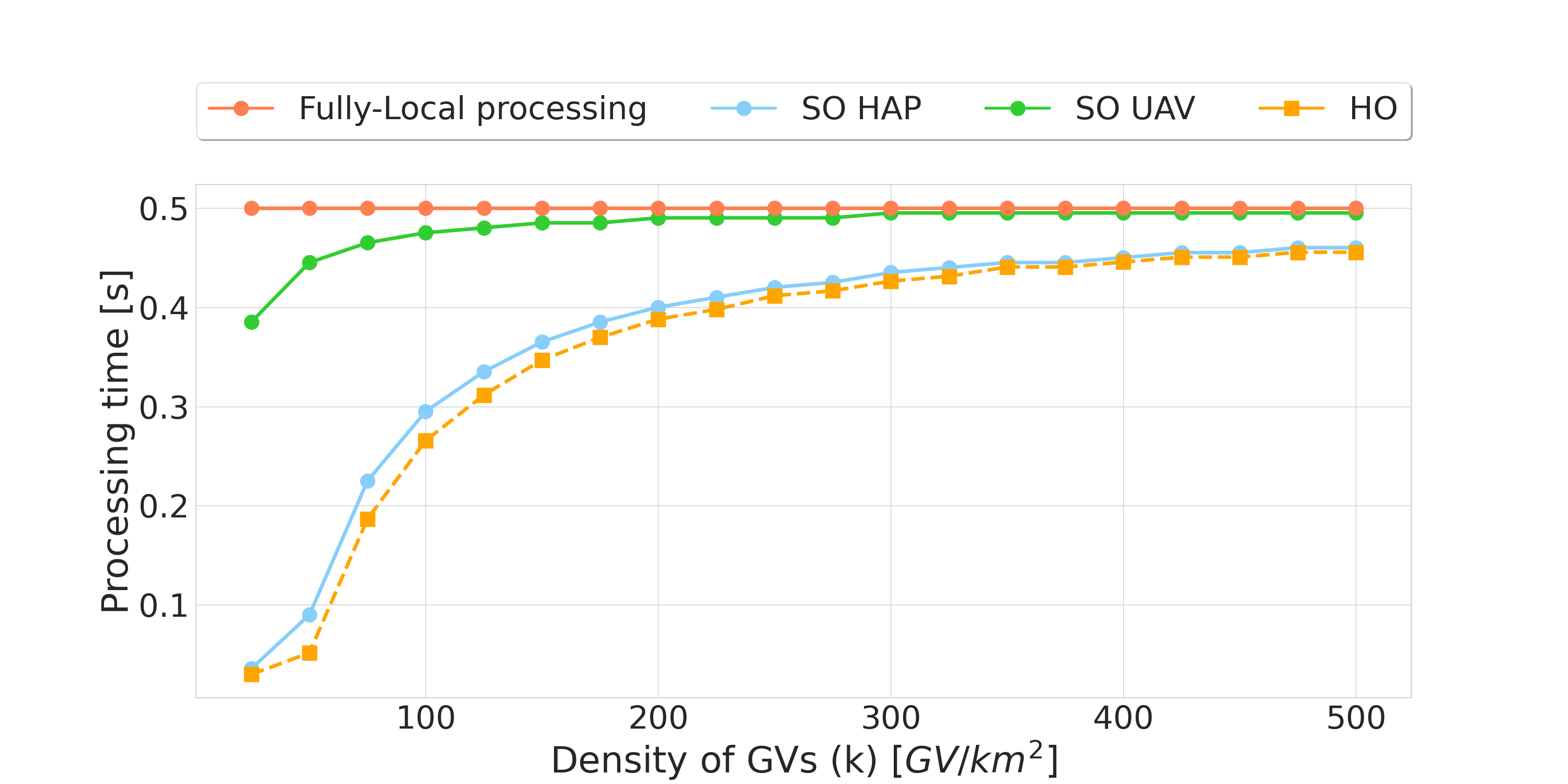}}\\
		\subfloat[][$C_{\rm GV}= 1000$~GFLOP/s.]
		{	\label{fig:users_density_c}
			\includegraphics[width=.87\columnwidth]{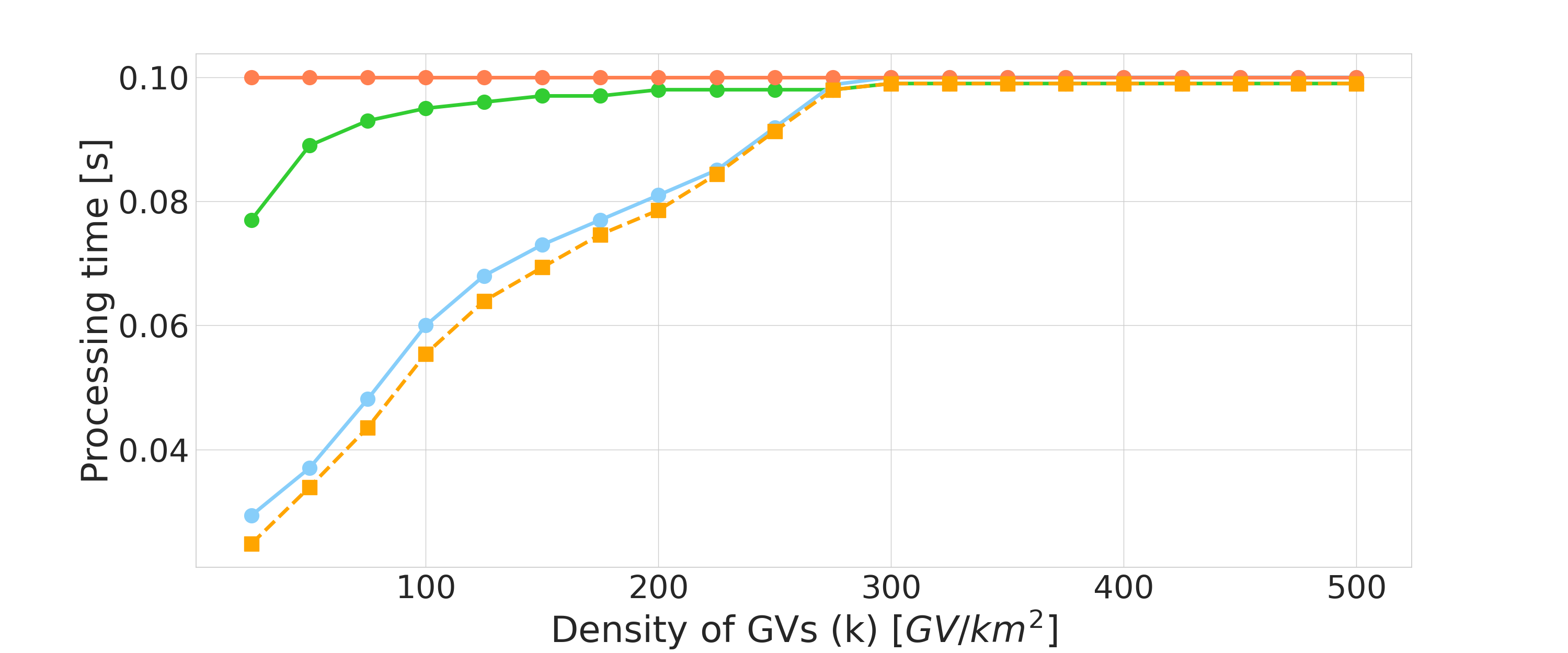}} 
		\caption{Offloading factor for different offloading schemes, vs. $C_{\rm GV}$ and~$k$.\vspace{-0.6cm}}
		\label{fig:users_density_offloading}
	\end{figure}

	In Figs.~\ref{fig:users_density_time} and \ref{fig:users_density_offloading} we evaluate the minimum processing time (and relative optimal offloading factor) for different offloading schemes, as a function of the density of GVs $k$ and the on-board computational capacity $C_{\rm GV}$.
	In Fig.~\ref{fig:users_density_time} each group of bar plots represents, respectively, standalone UAV offloading (SO UAV), standalone HAP offloading (SO HAP), and hybrid offloading (HO), respectively.
	We can see that, when $k=25$ (200), the optimal offloading factor for SO \gls{uav} is as low as 20\% (5\%), regardless of $C_{\rm GV}$. 
	This is due to the scarce computational capacity available to \glspl{uav}: despite the favorable channel, by offloading more tasks, the \gls{uav} queue would became unstable, i.e., $G(\eta_\text{UAV})> c_\text{UAV}$.
	For $C_{\rm GV}=200$ GFLOP/s (Fig.~\ref{fig:users_density_a}), the total processing time for SO~UAV is similar to that of fully-local computation, and is above 400 ms  even in sparsely deployed networks, which is incompatible with the requirements of most 5G/6G V2X applications. 
	For $C_{\rm GV}=1000$ GFLOP/s  (Fig.~\ref{fig:users_density_c}), the processing time becomes lower than $100$ ms, i.e., the frame rate of the sensors, which however involves higher hardware costs and energy consumption on board.

	In turn, SO HAP has access to more powerful computational resources and, despite the severe path loss, can support up to 95\% (70\%) of the computational load when $C_{\rm GV}=200$ ($1000$) GFLOP/s, for $k=25$.
	In this configuration the processing time is lower than 100 ms when $k<50$ even for $C_{\rm GV}=200$ GFLOP/s.
 	Notice that more populated scenarios (e.g., $k>200$) may overload the available channel bandwidth, and result in very long transmission~delays: in this case fully-local processing is the most desirable~option.
	
	As expected, the best scenario is the HO scheme, in which \gls{uav} and \gls{hap} nodes cooperate together to reduce the processing time.
	Nevertheless, the performance is similar to that of the SO HAP scheme, while in turn resulting in more complex and expensive network management.

	\begin{figure}[]
		\centering
		\subfloat[][$C_{\rm HAP}=3500$~GFLOP/s.]
		{	\label{fig:tx_bits_b}
			\includegraphics[width=.87\columnwidth]{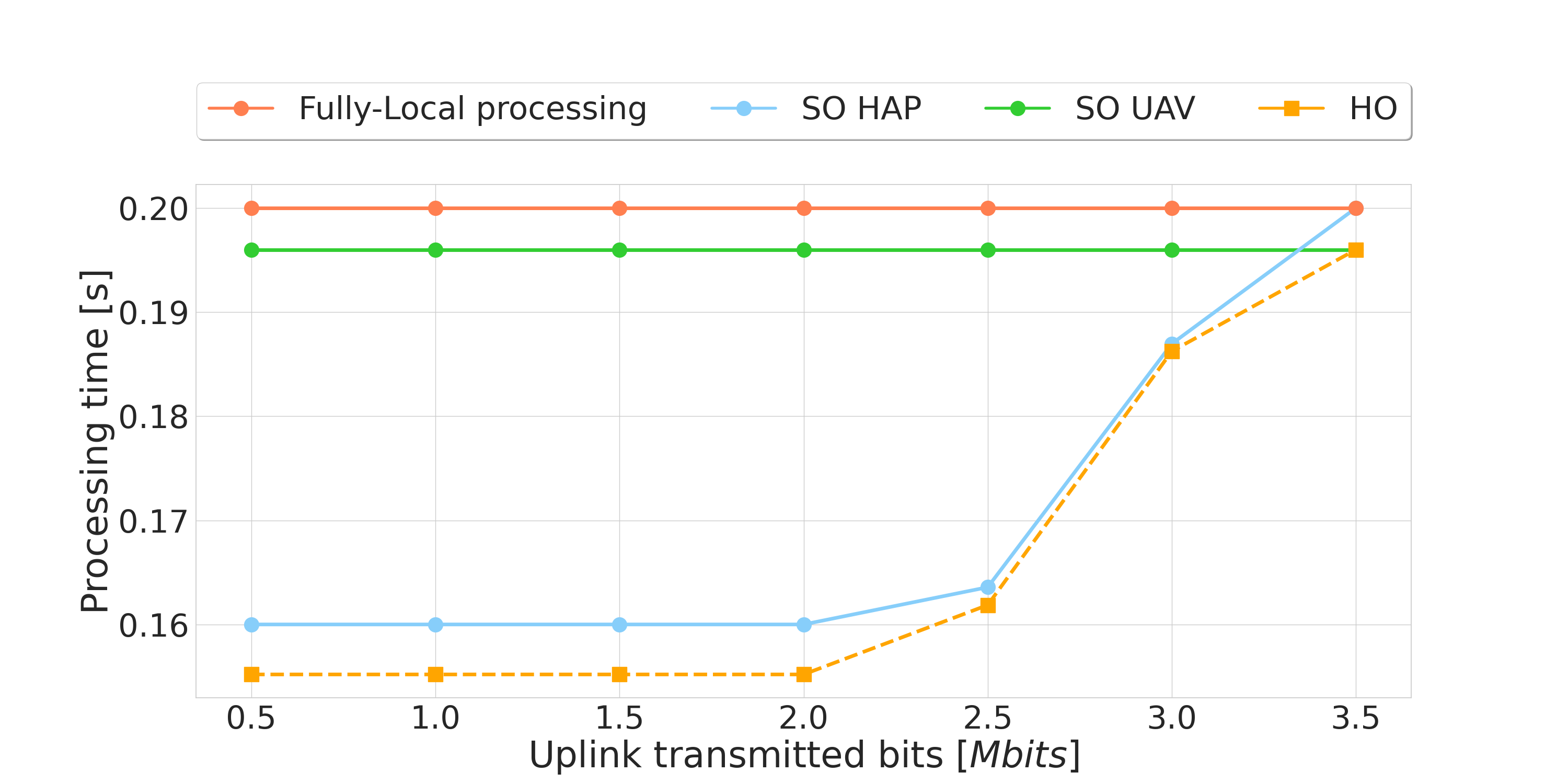}} \\
		\subfloat[][$C_{\rm HAP}=5000$~GFLOP/s.]
		{	\label{fig:tx_bits_c}
			\includegraphics[width=.87\columnwidth]{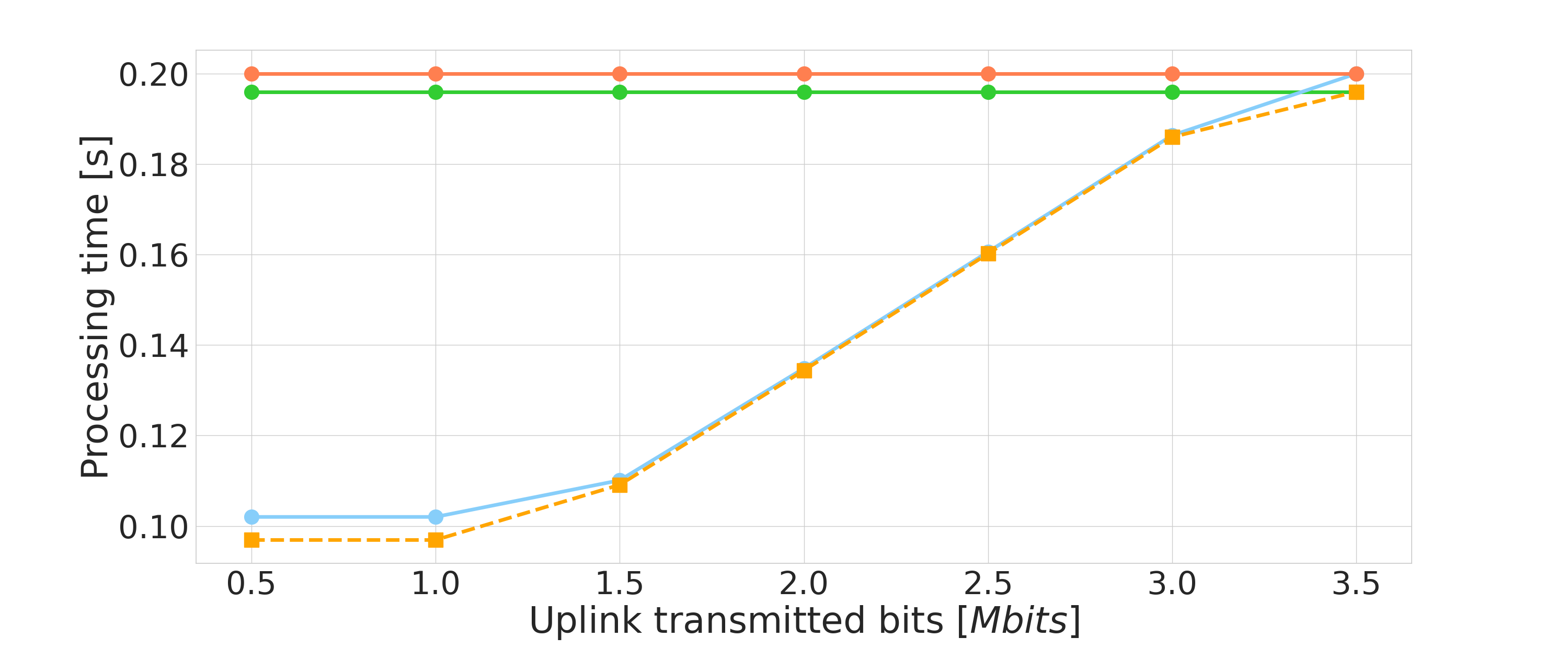}}		
		\caption{Total processing time for different offloading schemes and $k=200$, vs. $C_{\rm HAP}$ and $n_{\rm UL}$.\vspace{-0.6cm}}
		\label{fig:tx_bits}
	\end{figure}

	\subsubsection{Sensor's perception size}
	In this second set of simulations we evaluate the impact of the sensor's perception size $n_{\rm UL}$, which affects the transmission time for offloading described in Eq.~\eqref{eq:txdelays} and, thus, the total processing time.
		
	In Fig.~\ref{fig:tx_bits_b} we recognize two regimes for both the SO~HAP and HO configurations when $k=200$.
	When $n_{\rm UL}<2.5$~Mb, the total processing time is constant with $n_{\rm UL}$. 
	In this region, at most 20\% of the computational load can be offloaded to the HAP platform to maintain the system stable (Fig.~\ref{fig:users_density_time}), which means that  the remaining 80\% of the computational tasks have to be processed on board: this requires around 160~ms, regardless of the size of the perception data.
	When $n_{\rm UL}>2.5$~Mb, instead, the total processing time grows linearly with $n_{\rm UL}$.
	In this region the transmission time for offloading becomes so large that it exceeds the time for local processing, which then becomes more desirable. 
	On the other hand, in the SO UAV configuration, the impact of $n_{\rm UL}$ is negligible: the very low path loss experienced in the UAV channel results in a very low transmission delay, regardless of the size of the data. 
	The bottleneck is given by the UAV queue: given that less than 5\% of the computational load can be offloaded to the UAV platform  (Fig.~\ref{fig:users_density_time}), the processing time is similar to the case of local processing, and is around $200$~ms.

	From Fig.~\ref{fig:tx_bits_b} it appears clear that, even in the most convenient architecture, the total processing time is always above 100 ms (our target requirement for real-time operations). Better results can be obtained by considering more powerful offloading platforms. For example, in Fig.~\ref{fig:tx_bits_c} we see that, with a \gls{hap} with $c_{\rm HAP}=20$ servers, each of which has a computational capacity of $C_{\rm HAP}=5000$~GFLOP/s, the total processing time is maintained below the threshold of 100~ms as long as $n_\text{UL}<1.2$~Mb.

	\begin{figure}[t!]
		\centering
		{	\includegraphics[width=.66\columnwidth]{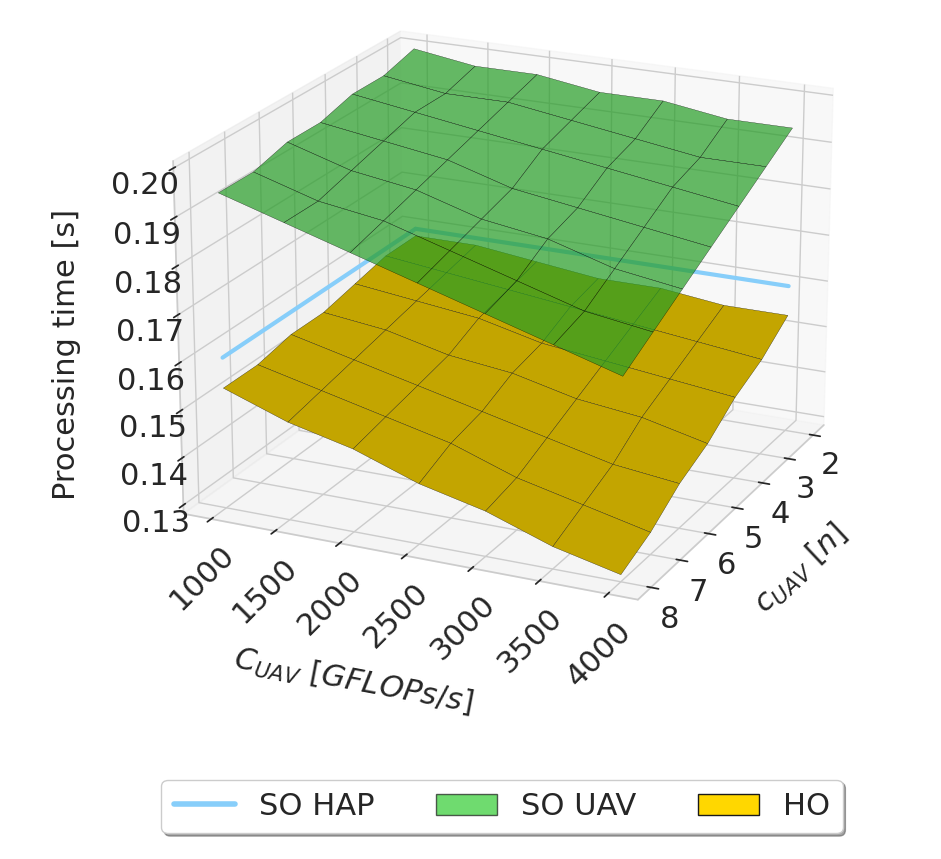}} 	
		{	\includegraphics[width=.66\columnwidth]{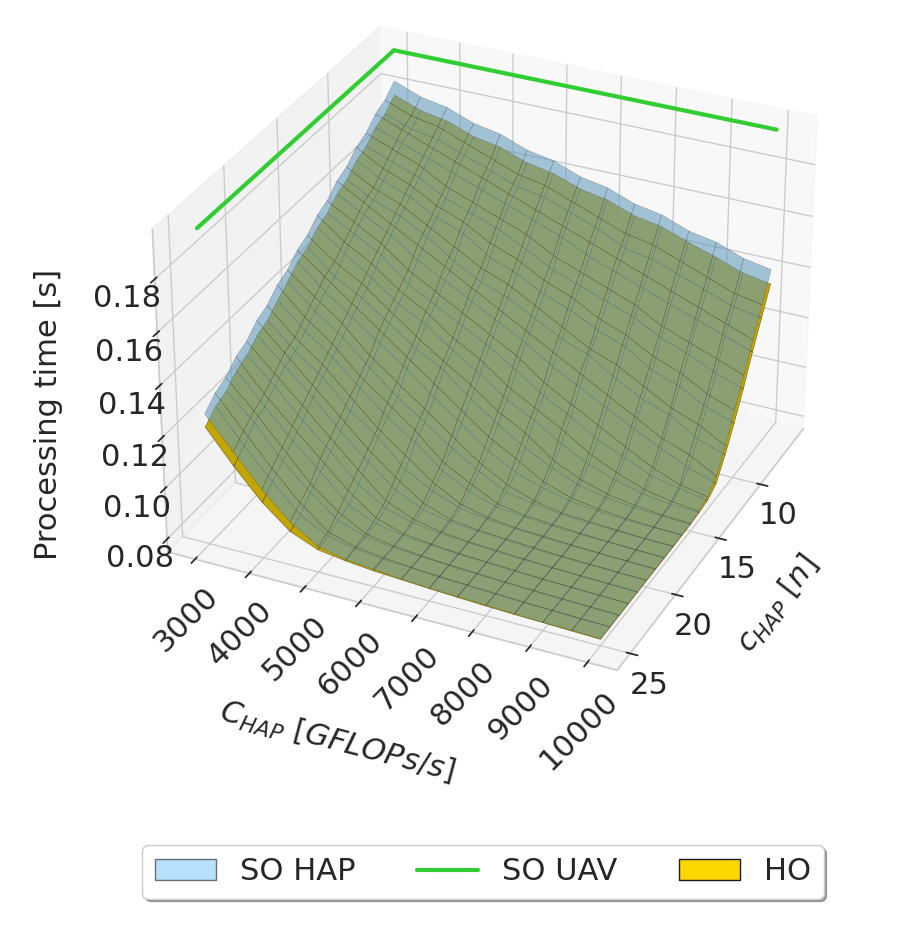}} 	
		\caption{Total processing time for different offloading schemes, vs. the UAV (up) and HAP (down) computational capacity, for $k=200$.\vspace{-0.5cm}}
		\label{fig:cap_proc_nt}
	\end{figure}

	\subsubsection{UAV/HAP computational capacity}
	As discussed above, the performance of the system is constrained by the capacity of the queues at the NTN servers, which can be expanded by increasing the computing power. 
	This is validated in Fig.~\ref{fig:cap_proc_nt} considering a scenario with $k=200$ GV/km$^2$, where we plot the processing time as a function of the computational capacity of UAVs and HAPs.
	For UAVs, $C_{\rm UAV}$ ranges between 1000 and 4000~GFLOP/s, whereas $c_{\rm UAV}$ varies between 2 and 8. 
	For HAPs, $C_{\rm HAP}$ ranges between 3000 and 10\,000~GFLOP/s, whereas $c_{\rm HAP}$ varies between 6 and~25.
		
	From Fig.~\ref{fig:cap_proc_nt} (up) we can see that, as the UAV becomes more powerful, the probability of offloading grows accordingly, which has the benefit to decrease the total processing time by up to 70\% compared to the baseline SO UAV results in Fig.~\ref{fig:users_density_a}.
	However, even considering HO, the processing time is still above the threshold of 100 ms, an indication that increasing the UAV's capacity alone is not enough to support V2X operations at the frame rate of the sensors.
	To achieve this objective, it is necessary to increase $C_{\rm HAP}$, as illustrated in Fig.~\ref{fig:cap_proc_nt} (down): in this case, the processing time is below 100 ms as long as $C_{\rm HAP}>5000$ GFLOP/s and $c_{\rm HAP}>15$.

	\section{Conclusions and Future Works} 
	\label{sec:conclusions_and_future_works}
	In this paper we shed light on the potential of NTN-assisted VEC for 6G systems.
	To do so, we considered a V2X scenario at a road intersection in which vehicles can decide to either process data on board, or offload a fraction of their computational load to UAVs and/or HAPs. Then, we formalized an optimization problem to find the optimal dimensioning of the computational capacity of GVs, UAVs, and HAPs to minimize the total processing time.
	From our simulations, we demonstrated that HAP-assisted VEC can reduce the processing time compared to fully-local processing, despite the transmission time required for offloading data to/from edge servers.
	We also showed that hybrid NTN offloading can further improve the processing performance, while incurring non-negligible hardware and computing costs.
	
	As part of our future work we will consider end-to-end system-level simulations.
	Moreover, we will extend our optimization problem to incorporate the impact of the energy consumption on the offloading performance.
	
	\section*{Acknowledgments} 
	Part of this work was supported by the US Army Research Office under Grant no. W911NF1910232. 
	
	\bibliographystyle{IEEEtran}
	\bibliography{bibliography.bib}
	
	\end{document}